%% file: Main.tex
\documentclass[journal]{IEEEtran}

\usepackage{cite}

\ifCLASSINFOpdf
\else
\fi
\usepackage[cmex10]{amsmath,mathtools}
\usepackage{amsfonts,amssymb}
\usepackage{amsfonts}

\usepackage{tikz}
\usetikzlibrary{shapes.geometric, arrows}

\tikzstyle{rect} = [rectangle, minimum width=2cm, minimum height=1cm, text centered, draw=black, fill=blue!3]
\tikzstyle{arrow} = [thick,->,>=stealth]

\usepackage{algorithm}
\usepackage{algcompatible} 
\usepackage{algpseudocode}

\usepackage{setspace}

\usepackage[latin1]{inputenc} 

\usepackage{graphicx}
\usepackage{epstopdf}
\usepackage[process=auto]{pstool} 
\usepackage{psfrag}

\usepackage{wrapfig}

\usepackage{relsize}

\usepackage[bookmarks=false,implicit=false,hidelinks]{hyperref}

\newcommand{\algorithmfootnote}[2][\footnotesize]{%
  \let\old@algocf@finish\@algocf@finish
  \def\@algocf@finish{\old@algocf@finish
    \leavevmode\rlap{\begin{minipage}{\linewidth}
    #1#2
    \end{minipage}}%
  }%
}

\usepackage{setspace}

\usepackage{bbm}

\newcommand{\that}[1]{\skew{4}\hat{\boldsymbol{\theta}}}

\begin{document}

\title{Sensor Fusion for Magneto-Inductive Navigation}
\author{Johan Wahlstr\"{o}m, Manon Kok, Pedro Porto Buarque de Gusm\~ao, \\Traian E. Abrudan, Niki Trigoni, and Andrew Markham
\thanks{This research has been financially supported by the National Institute of Standards and Technology (NIST) via the grant \emph{Pervasive, Accurate, and Reliable Location-based Services for Emergency Responders} (Federal Grant: 70NANB17H185).}
\thanks{J. Wahlstr\"om, Pedro Porto Buarque de Gusm\~ao, N. Trigoni, and A.Markham are with the Department of Computer Science, University of Oxford, Oxford, UK (\{johan.wahlstrom, pedro.gusmao, niki.trigoni, andrew.markham\}@cs.ox.ac.uk).}
\thanks{Manon Kok is with Delft University of Technology, Delft, the Netherlands (e-mail: m.kok-1@tudelft.nl).}
\thanks{Traian E. Abrudan is with Nokia Bell Labs, Dublin, Ireland (e-mail: traian.abrudan@nokia-bell-labs.com). Traian's contribution was completed while he was with the Department of Computer Science, University of Oxford.}
}


%


\maketitle

\begin{abstract}
Magneto-inductive navigation is an inexpensive and easily deployable solution to many of today's navigation problems. By utilizing very low frequency magnetic fields, magneto-inductive technology circumvents the problems with attenuation and multipath that often plague competing modalities. Using triaxial transmitter and receiver coils, it is possible to compute position and orientation estimates in three dimensions. However, in many situations, additional information is available that constrains the set of possible solutions. For example, the receiver may be known to be coplanar with the transmitter, or orientation information may be available from inertial sensors. We employ a maximum a posteriori estimator to fuse magneto-inductive signals with such complementary information. Further, we derive the Cram\'er-Rao bound for the position estimates and investigate the problem of detecting distortions caused by ferrous material. The performance of the estimator is compared to the Cram\'er-Rao bound and a state-of-the-art estimator using both simulations and real-world data. By fusing magneto-inductive signals with accelerometer measurements, the median position error is reduced almost by a factor of two. 


\end{abstract}

%
\IEEEpeerreviewmaketitle

\section{introduction}

Indoor navigation technology based on very low frequency (kHz) magneto-inductive (MI) fields has several attractive characteristics. In particular, MI fields penetrate soil, concrete, and rock with negligible attenuation, and are not subject to multipath or shadow fading \cite{Vuran2010,Sheinker2013,Arumugam2012}. Moreover, the technology is inexpensive, has low power consumption, and enables 3-D position and orientation estimation using a single transmitter-receiver pair \cite{Abrudan2015,Abrudan2016,Abrudan2016b,Abrudan2017}. 
Studied applications include structural health monitoring \cite{Kypris2017}, underground pipeline monitoring \cite{Sun2011}, animal tracking \cite{Markham2010}, and emergency rescue \cite{Wei2018}. The small setup effort, combined with the favorable characteristics with respect to attenuation and multipath, also makes magneto-inductive navigation particularly suitable for firefighter positioning \cite{Ferreira2017}. Recent studies within MI navigation have focused on several overlapping categories of estimation and detection problems. These include position and range estimation, orientation estimation, detection and estimation of distortions, and sensor fusion.

\begin{figure}[t]
	\def\svgwidth{0.8\columnwidth}
	\hspace*{5mm}
	\scalebox{1.1}{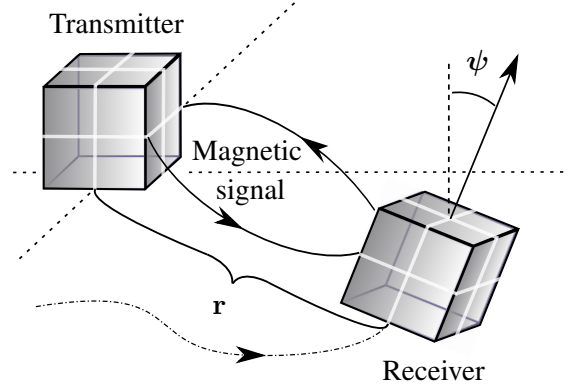}
	\vspace*{-2mm}
	\caption{A magneto-inductive field is generated using a stationary transmitter. Based on the transmitted signal, a mobile receiver can estimate its position $\mathbf{r}$ and orientation $\boldsymbol{\psi}$ with respect to the transmitter.}
	\label{fig_illustration}
	\vspace*{-0mm}
\end{figure}

\subsection{Current Challenges within MI Navigation}

As illustrated in Fig. \ref{fig_illustration}, the position and orientation estimates are based on a modulated MI signal, sent from a triaxial transmitter to a triaxial receiver. After demodulating the signal \cite{Markham2012}, information on the navigation state can be extracted based on the dipole field model. In \cite{Arumugam2012,Arumugam2012b,Arumugam2014}, the position and orientation estimates were computed using least squares. In \cite{Dumphart2017}, the maximum likelihood estimates were found by alternating between optimizing for position and orientation (only the pitch and yaw angles were considered). In \cite{Abrudan2015} and \cite{Abrudan2016}, the channel matrix was estimated using least squares, and the range was estimated based on the received signal strength indicator (RSSI). After eliminating the orientation from the dipole field model using algebraic manipulations, the model was inverted to produce estimates of the position direction. Finally, the orientation was estimated by using the estimated position and channel matrix to invert the dipole field model.

The rapid path loss of the dipole field model is a double-edged sword \cite{Sun2010,Pasku2017a,Pasku2017b,Huang2018,Hehn}. It enables highly accurate navigation estimates at short distances, but also puts strict restrictions on the practical navigation range. Another challenge imposed by the dipole field model is the hemispherical ambiguity, i.e., that the sign of the position is not identifiable when using measurements from one receiver-transmitter pair alone. This ambiguity can be resolved by using multiple transmitters or by considering additional information from e.g., maps or inertial measurements \cite{Wei2018}. 

The fact that the position and orientation may change whilst receiving a transmission is an additional challenge. In \cite{Abrudan2015}, the MI signal was rotated based on gyroscope measurements to compensate for changes in the orientation during transmission. The same study proposed several tests for assessing the reliability of position and orientation estimates. If the tests indicated that the estimates were unreliable (due to e.g., inadequate rotational compensation), the navigation system reverted to estimating range alone based on the total received power. 

MI estimates may also be adversely affected by ferrous materials causing magnetic distortions in the environment. In \cite{Kypris2015} and \cite{Kypris2016}, image theory was used to model the magnetic field above a floor with metallic reinforcement bars. Specifically, the floor was modeled as a perfectly conducting half-space, and the magnetic field model was obtained by placing an image dipole underneath the floor (mirroring the position of the transmitter). 

Fusion of MI and inertial measurements have been studied in several contexts. In \cite{Abrudan2015}, inertial sensors were used to not only compensate for rotational changes during transmission, but also to bridge gaps between MI updates and increase the update rate of the navigation solution. Similarly, in \cite{Wei2018}, MI and inertial measurements were fused by using a simultaneous localization and mapping (SLAM) algorithm that exploited local distortions in the magnetic field. Fusion of MI signals from multiple transmitters has been studied in \cite{Abrudan2016}. Although the concept of MI navigation has been around for many decades, no one has yet taken a hard look at the associated general problem of low-level information fusion.

\subsection{Related Navigation Technologies}

There are several ways to utilize measurements of the ambient magnetic field for navigation purposes. If local magnetic disturbances are negligible, a magnetometer triad can be used to find the direction of the magnetic North \cite{Groves2008}. In the presence of stationary and non-uniform magnetic distortions, the magnetic field can instead be used for positioning or odometry. Magnetic positioning requires a magnetic field map where local anomalies are associated with a given set of positions \cite{Kok2018}. Although large spatial gradients enable a high positioning accuracy, they also put higher demands on the resolution of the map. Further, note that magnetic positioning is limited by the fact that the fingerprint consists of at most three parameters. In magnetic odometry, an array of magnetometers is used to obtain a local approximation of the magnetic field \cite{Chesneau2016,Caruso2017,Caruso2017b,Caruso2018}. By comparing the measurements at two sequential sampling instances, it is possible to estimate the translational and angular velocity. Thus, magnetic odometry does not require a magnetic field map, and differs from magnetic positioning both in terms of the required hardware and the produced navigation estimates. However, the translational velocity can only be estimated in the body frame, and therefore, knowledge about the array orientation is needed to determine the velocity in a fixed navigation frame. Comparing these technologies with magneto-inductive navigation, we conclude that only the latter can produce position and orientation estimates without mapping the magnetic field. Further, by generating a magneto-inductive field, it is possible to use the magnetic field for localisation also in environments where the ambient magnetic field is homogeneous and other magnetic localisation methods would not work. Another way to compute position and orientation estimates from magnetic field measurements is by tracking a permanent magnet \cite{Skog2018}. However, this puts limitations on the dipole moment, which, in turn, generally will lead to a reduced estimation performance.

%
%


\subsection{Contributions and Findings}

In this paper, MI navigation is approached from a traditional estimation perspective. In particular, 
\begin{itemize}
\item We propose a maximum a posteriori (MAP) estimator for the problem of fusing MI measurements and eventual supplementary information to estimate position and orientation. As opposed to previous estimators \cite{Abrudan2015,Abrudan2016}, the proposed estimator is derived based on a conventional optimality criterion and simultaneously computes position and orientation estimates in a one-stage procedure. Moreover, the proposed estimator provides a general framework for tight fusion (all available information is fused before computing any navigation estimates) of MI signals and additional information obtained from e.g., previous MI estimates, motion models, maps, or measurements from other sensors. 
\item We develop a method for detection of magnetic distortions and inadequate compensation for changes in the navigation state during transmission. As opposed to previous detectors \cite{Abrudan2015}, the proposed detector uses a probabilistic model for the studied test statistic. This enables straightforward computation of p-values. 
\item Assuming perfect knowledge of the orientation, we derive closed-form Cram\'er-Rao bounds (CRBs) for the position and range estimates.
\end{itemize}
Among our findings, we wish to highlight the following: 
\begin{itemize}
\item Numerical comparison reveals that the position and orientation estimator presented in \cite{Abrudan2015} and \cite{Abrudan2016} (and outlined in Appendix A) is the MAP estimator with an uninformative prior, i.e., the maximum likelihood (ML) estimator. 
\item The sum of the position Fisher information along all spatial directions is independent of the direction of the position vector and the orientation. However, given some fixed range, the position Cram\'er-Rao bound is shown to be lower in the spatial directions where the distance between the transmitter and the receiver is large. In practice, this means that when moving in an area where the magnitude of the altitude is small in comparison to the magnitude of the horizontal position (such as when walking in a one-storey building), the position estimates can be expected to be worse in the altitude direction than in the horizontal plane. This aligns well with the experimental results in this paper and in \cite{Abrudan2015}, where it was stated that ``in all indoor environments we have tested, the dominant component of the positioning error is along the z-axis''.
\item Through simulations, it is established that prior information on the orientation have the potential of reducing the root-mean-square error (RMSE) of the position estimates by more than a factor of two.
\item In an experiment where magneto-inductive signals are fused with measurements from low-cost accelerometers, the median position error is reduced almost by a factor of two. Thus, this demonstrates that measurements from inertial sensors can significantly improve the performance of MI navigation, with negligible increases in cost and setup complexity.
\end{itemize}


\emph{The data and the code used in the experiments are available at \href{https://www.cs.ox.ac.uk/people/johan.wahlstrom/}{\textcolor{blue}{https://www.cs.ox.ac.uk/people/johan.wahlstrom/}}}.

\section{estimation}
\label{section_estimation}

This section describes the signal model and the MAP estimator for one transmitter-receiver pair. In addition, we give examples of position and orientation priors. The estimator is derived under the assumption that satisfactory signal demodulation has been performed and that the navigation state is constant during the transmission of a frame (the method for rotation stabilization presented in \cite{Abrudan2015} is outlined in Appendix A). The extension to the case with multiple transmitters is straightforward. 

\subsection{Signal Model}

The $k$th 3-D MI measurement obtained by the receiver in a given transmission can be described as
\begin{equation}
\label{eq_meas_model}
\mathbf{y}_k=\mathbf{h}_{\mathbf{m}_k}(\mathbf{x})+\mathbf{e}_k
\end{equation}
where the dipole field is
\begin{equation}
\label{eq_dipole_field}
\mathbf{h}_{\mathbf{m}}(\mathbf{x})\overset{_\Delta}{=}c\cdot\frac{\mathbf{R}(\boldsymbol{\psi})}{\Vert\mathbf{r}\Vert^3}\bigg(\frac{3\mathbf{r}\mathbf{r}^\intercal}{\Vert\mathbf{r}\Vert^2}-\mathbf{I}_3\bigg)\mathbf{m}.
\end{equation}
The navigation state  $\mathbf{x}\overset{_\Delta}{=}[\mathbf{r}^\intercal\;\boldsymbol{\psi}^\intercal]^\intercal$ collects the position $\mathbf{r}$ and orientation $\boldsymbol{\psi}$ (represented using Euler angles) of the receiver with respect to the transmitter frame. Further,
$\mathbf{e}_k$ is zero-mean Gaussian\footnote{The assumption of zero-mean Gaussian measurement noise has previously been verified in \cite{Abrudan2015}.} measurement noise with covariance $\mathbf{P}$, $c$ is a known scale factor given by the properties of the receiver-transmitter pair, $\mathbf{R}$ is the three-dimensional rotation matrix describing rotations from the transmitter frame to the receiver frame, $\mathbf{I}_n$ is the identity matrix of dimension $n$, $\mathbf{m}_k$ is the known magnetic moment transmitted at sampling instance $k$, and $\Vert\cdot\Vert$ is the Euclidean norm. Refer to \cite{Abrudan2015} for further details on the model.
Now, the problem at hand is to estimate the navigation state $\mathbf{x}$ from the MI measurements $\mathbf{y}\overset{_\Delta}{=}\{\mathbf{y}_k\}_{k=1}^N$.  

\subsection{Maximum-a-Posteriori Estimation}

The MAP estimate of $\mathbf{x}$ is
\begin{align}
\begin{split}
\hat{\mathbf{x}}&=\underset{\mathbf{x}}{\arg\max}\;p(\mathbf{x}|\mathbf{y})\\
&=\underset{\mathbf{x}}{\arg\max}\,\ln p(\mathbf{y}|\mathbf{x}) + \ln p(\mathbf{x}). 
\end{split}
\end{align}
The posterior distribution $p(\mathbf{x}|\mathbf{y})$ includes information from both the likelihood function $p(\mathbf{y}|\mathbf{x})$ and the prior $p(\mathbf{x})$. The likelihood function depends on the MI measurements (it is a Gaussian distribution with a mean given by \eqref{eq_dipole_field} and the covariance $\mathbf{P}$), while the prior $p(\mathbf{x})$ can be based on eventual side-information obtained from e.g., previous MI estimates, motion models, maps, or measurements from other sensors. Iterative estimation algorithms can be initialized by using the estimates produced by the estimation method outlined in Appendix A. 

\subsection{The Uninformative Prior}

With an uninformative prior $p(\mathbf{x})\propto 1$, the MAP estimate is equivalent to the ML estimate 
\begin{align}
\begin{split}
\hat{\mathbf{x}}&=\underset{\mathbf{x}}{\arg\max}\,\ln p(\mathbf{y}|\mathbf{x})\\
&=\underset{\mathbf{x}}{\arg\min}\,\sum_{k=1}^N\Vert\mathbf{y}_k-\mathbf{h}_{\mathbf{m}_k}(\mathbf{x})\Vert_{\mathbf{P}^{-1}}^2
\end{split}
\end{align}
where $\Vert\mathbf{u}\Vert_{\mathbf{A}}^2\overset{_\Delta}{=}\mathbf{u}^\intercal\mathbf{A}\mathbf{u}$. This is a nonlinear least-squares problem which can be solved by standard methods, such as the Gauss-Newton algorithm, the Levenberg-Marquardt algorithm, or trust-region methods \cite{Nocedal2006}. The estimator presented in \cite{Abrudan2015} and \cite{Abrudan2016} is equivalent to the ML estimator, and will therefore not be presented as a separate estimator in the performance evaluations.

\subsection{Position Priors}

Given the ubiquity of Gaussian distributions, we consider a Gaussian position prior with mean $\boldsymbol{\mu}_r$ and covariance $\boldsymbol{\Sigma}_r$. Assuming there is no prior orientation information, the MAP estimate then becomes
\begin{align}
\hat{\mathbf{x}}=\underset{\mathbf{x}}{\arg\min}\,&\sum_{k=1}^N\Vert\mathbf{y}_k-\mathbf{h}_{\mathbf{m}_k}(\mathbf{x})\Vert_ {\mathbf{P}^{-1}}^2\\\nonumber
&+\Vert\boldsymbol{\mu}_r-\mathbf{g}_r(\mathbf{x})\Vert_{\boldsymbol{\Sigma}_r^{-1}}^2
\end{align}
where $\mathbf{g}_r(\mathbf{x})\overset{_\Delta}{=}\mathbf{r}$. Once again, the MAP estimate is found by solving a nonlinear least-squares problem. More sophisticated position priors may be obtained when using information from e.g., WiFi fingerprinting or maps. 

\subsection{Orientation Priors}

Orientation priors may also be based on Gaussian distributions. For example, consider a Gaussian prior on the Euler angles\footnote{The use of flat distributions (such as the Gaussian) to model the distribution of circular parameters (such as the Euler angles) does typically not cause any problems as long as the uncertainty is small in comparison to the range of the parameter space \cite{Gerwe2003}.} with mean $\boldsymbol{\mu}_\psi$ and covariance $\boldsymbol{\Sigma}_{\psi}$. 
Defining $\mathbf{g}_\psi(\mathbf{x})\overset{_\Delta}{=}\boldsymbol{\psi}$ and assuming an uninformative position prior then gives the MAP estimate
\begin{align}
\hat{\mathbf{x}}=\underset{\mathbf{x}}{\arg\min}\,&\sum_{k=1}^N\Vert\mathbf{y}_k-\mathbf{h}_{\mathbf{m}_k}(\mathbf{x})\Vert_ {\mathbf{P}^{-1}}^2\\\nonumber
&+\Vert\boldsymbol{\mu}_\psi-\mathbf{g}_\psi(\mathbf{x})\Vert_{\boldsymbol{\Sigma}_\psi^{-1}}^2
\end{align}
which, yet again, defines a nonlinear least-squares problem. 

Although there are many alternative orientation distributions, the Gaussian distribution considered above is practical when the prior only provides information along some dimensions of the orientation space. For example, at standstill\footnote{Another way of extracting orientation information from inertial sensors is to use foot-mounted sensors. By using zero-velocity updates, foot-mounted inertial sensors can provide highly accurate roll and pitch estimates also when in motion \cite{Wahlstrom2019}.}, accelerometers provide information on the roll and pitch angles (but not the yaw angle) with respect to a tangent frame. In addition, the considered prior enables straightforward computations of the MAP estimate and the CRB. Refer to \cite{Kok2017} for a review of probabilistic orientation modeling.

\section{fault detection}
\label{section_detection}

Using standard probability theory, we have that
\begin{equation}
T(\mathbf{x})\overset{_\Delta}{=} \sum_{k=1}^N\Vert\mathbf{y}_k-\mathbf{h}_{\mathbf{m}_k}(\mathbf{x})\Vert_{\mathbf{P}^{-1}}^2\sim\chi^2_{3N}
\end{equation}
where $\chi^2_{\kappa}$ denotes a chi-squared distribution with $\kappa$ degrees of freedom. Hence, a natural way to test the validity of the model is to perform a chi-squared test with $T(\hat{\mathbf{x}})$ as the test statistic. A rejection of the null hypothesis could be the result of magnetic distortions or a deficient compensation for changes in the navigation state during transmission. Note, however, that the proposed chi-squared test cannot be used to detect any given magnetic distortion. For example, consider distortions that can be described as rotations of the measurements. Such distortions will just rotate the orientation estimates, without increasing the value of the test statistic. 

\section{cram\'er-rao bound}
\label{section_CRB}

The CRB provides a lower bound on the mean square error of an estimator. It is often used to characterize an estimation problem in terms of its underlying parameters. Thus, by studying the CRB, one may discover inherent limitations of the problem at hand, or gain insight into how experiments can be designed to increase estimation performance. Moreover, the CRB can also be used to evaluate the performance of estimators. If the mean-square error of an estimator is in the vicinity of the bound, the estimator is typically considered to be adequate. 

For brevity and tractability, we will focus on the scenario where we have perfect knowledge of the orientation. That is, when we want to estimate the position $\mathbf{r}$ from the MI measurements $\{\mathbf{y}_k\}_{k=1}^N$ with the orientation $\boldsymbol{\psi}$ considered to be known. Since we are merely adding information to the original problem, this will still result in a lower bound for the RMSE of the position estimates in the scenario with imperfect knowledge of the orientation. CRBs for the case with multiple transmitters can be computed by using the additive property of the Fisher information.

\subsection{Cram\'er-Rao Bound for Position Estimates}

Consider the case when $\boldsymbol{\psi}$ is known, $\mathbf{m}_{1+3k}=[\hspace*{0.2mm}m\,0\,0\hspace*{0.2mm}]^\intercal$, $\mathbf{m}_{2+3k}=[\hspace*{0.2mm}0\,m\,0\hspace*{0.2mm}]^\intercal$, and $\mathbf{m}_{3+3k}=[\hspace*{0.2mm}0\,0\,m\hspace*{0.2mm}]^\intercal$ for all $k$, $N$ is equal to a multiple of 3, and $\mathbf{P}=\sigma^2\cdot\mathbf{I}_3$. The CRB for $\hat{\mathbf{r}}$ (not considering prior information) then becomes 
\begin{equation}
\mathrm{Var}([\hat{\mathbf{r}}]_i)\ge \mathcal{I}_{r_i}^{-1}
\end{equation}
where the Fisher information for $[\mathbf{r}]_i$ is
\begin{equation}
\label{eq_fish_inf}
\mathcal{I}_{r_i}=\frac{6Nc^2m^2}{\sigma^2\Vert\mathbf{r}\Vert^8}\bigg(1+2\frac{[\mathbf{r}]_i^2}{\Vert\mathbf{r}\Vert^2}\bigg).
\end{equation}
Here, $[\mathbf{r}]_i$ is the $i$th element of the position vector. A derivation of the Fisher information is provided in Appendix C.

The Fisher information in \eqref{eq_fish_inf} is factorized as a product of two factors. The first depends on the range $\Vert\mathbf{r}\Vert$, and the second depends on the relative distance $|[\mathbf{r}]_i|/\Vert\mathbf{r}\Vert$. Two noteworthy observations can be made from this:
\begin{itemize}
\item Given some fixed direction of the position vector $\mathbf{r}$, the CRB increases with the 8th power of the range $\Vert\mathbf{r}\Vert$.
\item The sum of the Fisher information along all three spatial directions equals
\begin{equation}
\sum_{i=1}^{3}\mathcal{I}_{r_i}=\frac{30Nc^2m^2}{\sigma^2\Vert\mathbf{r}\Vert^8} 
\end{equation}
and is independent of the direction of the position vector. According to \eqref{eq_fish_inf}, more Fisher information is allocated to spatial directions where the position distance $|[\mathbf{r}]_i|$ is large. Given some fixed range $\Vert\mathbf{r}\Vert$, the Fisher information for $[\mathbf{r}]_i$ is maximized when $|[\mathbf{r}]_i|=\Vert\mathbf{r}\Vert$.
\end{itemize}

\subsection{Cram\'er-Rao Bound for Range Estimates}

Consider the same setting as in the previous subsection. As shown in Appendix D, the CRB for $\Vert\hat{\mathbf{r}}\Vert$ then is 
\begin{equation}
\mathrm{Var}(\Vert\hat{\mathbf{r}}\Vert)\ge \mathcal{I}_{\Vert\mathbf{r}\Vert}^{-1}
\end{equation}
where
\begin{equation}
\label{eq_fish_inf_range}
\mathcal{I}_{\Vert\mathbf{r}\Vert}=\frac{18Nc^2m^2}{\sigma^2\Vert\mathbf{r}\Vert^8}.
\end{equation}
Comparing with \eqref{eq_fish_inf}, it can be seen that $\mathcal{I}_{r_i}\le\mathcal{I}_{\Vert\mathbf{r}\Vert}$. Thus, the CRB for the range estimates is always equal to or smaller than the CRB for the position estimates along any given spatial direction.  

\section{simulations}

Next, simulations are used to evaluate the estimation and detection algorithms proposed in Sections \ref{section_estimation} and \ref{section_detection}, respectively. Measurements were generated with $\mathbf{r}=[\hspace*{0.2mm}1\,1\,1\hspace*{0.2mm}]_{}^\intercal$, $\boldsymbol{\psi}=[\hspace*{0.2mm}0\,0\,0\hspace*{0.2mm}]_{}^\intercal$,
$c=1$, $\mathbf{m}_{1+3k}=[\hspace*{0.2mm}1\,0\,0\hspace*{0.2mm}]^\intercal$, $\mathbf{m}_{2+3k}=[\hspace*{0.2mm}0\,1\,0\hspace*{0.2mm}]^\intercal$, and $\mathbf{m}_{3+3k}=[\hspace*{0.2mm}0\,0\,1\hspace*{0.2mm}]^\intercal$ for $k=0,\dots,9$, $N=30$, and $\mathbf{P}=\sigma^2\cdot\mathbf{I}_3$ with
$\sigma=0.1$.

\subsection{Position and Orientation Estimation}

\begin{figure}[t]
\hspace*{-1.5mm}
\vspace*{3mm}
{
	\includegraphics{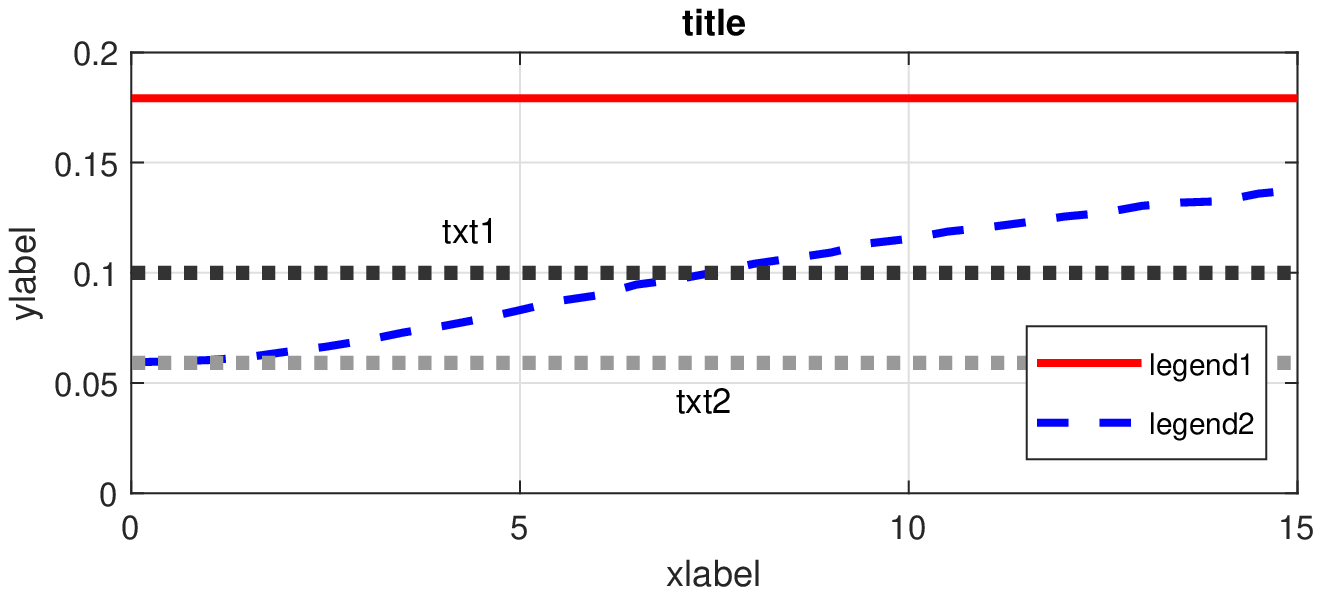}
\vspace*{3mm}
}
\hspace*{-2.2mm}
{
	\vspace*{-1.3mm}
	\includegraphics{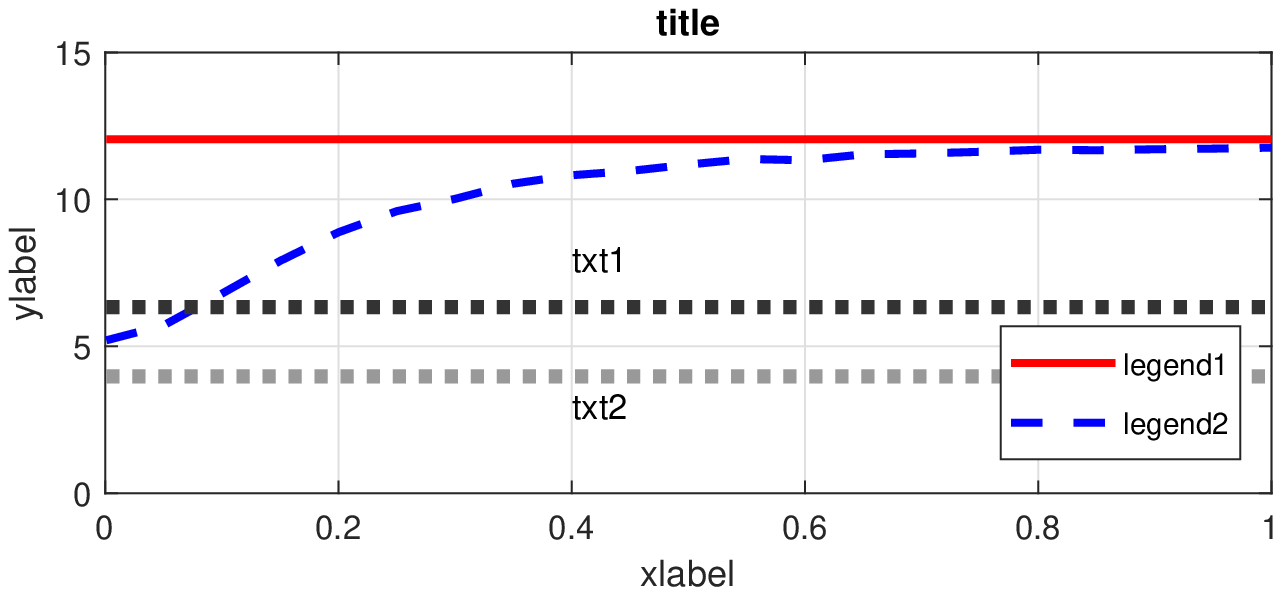}
\vspace*{0mm}
}
\vspace*{-8mm}
\caption{Accuracy of position and orientation estimates on simulated data.  \label{fig_simulations_accuracy}}
\vspace*{-3mm}
\end{figure}

The MAP and ML estimators presented in Section \ref{section_estimation} are now compared to CRBs. For ease of illustration, the Cramér-Rao inequality was reduced to the scalar
form $\mathbb{E}[\Vert\hat{\mathbf{r}}-\mathbf{r}\Vert^2]\ge\textstyle\mathrm{Tr}(\boldsymbol{\mathcal{I}}_\mathbf{r}\rule{-5pt}{0pt}^{-1})$, where $\boldsymbol{\mathcal{I}}_\mathbf{r}$ is the Fisher information matrix for $\mathbf{r}$. Hence, we compared the scalar RMSE $\sqrt{\mathbb{E}[\Vert\hat{\mathbf{r}}-\mathbf{r}\Vert^2]/3}$ with  $\sqrt{\smash{\mathrm{Tr}(\boldsymbol{\mathcal{I}}_\mathbf{r}\rule{-5pt}{0pt}^{-1})}/3}$ (and likewise for the orientation estimates). The hemispherical ambiguity was handled manually by switching the sign of $\hat{\mathbf{r}}$ whenever this decreased the distance between $\hat{\mathbf{r}}$ and $\mathbf{r}$.   

Fig. \ref{fig_simulations_accuracy} (a) shows the position RMSE of the ML and MAP estimators as a function of $\sigma_\psi$ when using an orientation prior with covariance $\boldsymbol{\Sigma}_{\psi}=\sigma_\psi^2\,\mathbf{I}_3$. The mean of the orientation prior was simulated from a Gaussian distribution with a mean equal to the true orientation and the covariance $\boldsymbol{\Sigma}_{\psi}$. The RMSEs were computed from $10^4$ such simulations. Further, Fig. \ref{fig_simulations_accuracy} (b) presents the corresponding orientation RMSEs obtained when using a position prior. As expected, the RMSE of the MAP estimator draws closer to that of the ML estimator as the amount of information in the prior diminishes. Moreover, the RMSE of the MAP estimator approaches the CRB with perfect prior information as the uncertainty of the prior decreases. For both the position and orientation estimates, perfect prior information on the orientation and position, respectively, gives a RMSE that is less than half that of the ML estimator. 

\begin{figure}[t]
	\vspace*{0mm}
	\hspace*{-1mm}
	\includegraphics{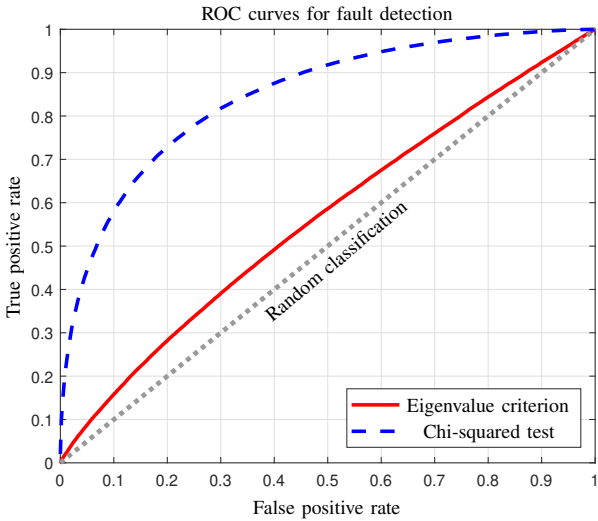}
	\vspace*{-6mm}
	\caption{Receiver operating characteristics on simulated data.}
	\vspace*{0mm}
	\label{fig_simulations_ROC}
\end{figure}

\subsection{Detection of Distortions}

In what follows, we benchmark two methods for fault detection: the chi-squared test described in Section \ref{section_detection} and the eigenvalue criterion presented in \cite{Abrudan2015} and outlined in Appendix B. The receiver operating characteristics (ROC) of the two detectors are displayed in Fig. \ref{fig_simulations_ROC}. Here, the false positive rate was computed as the percentage of rejections of the null hypothesis (i.e., that the dipole field model is correct and that the measurement errors are zero-mean Gaussians with covariance $\mathbf{P}$) when performing $10^5$ simulations based on dipole field model \eqref{eq_dipole_field}. Similarly, the true positive rate was computed as the percentage of null hypothesis rejections when simulating measurements from the same model but with a measurement error covariance of $2\cdot\mathbf{P}$ (the added noise was used to represent distortions). As can be seen from Fig. \ref{fig_simulations_ROC}, the chi-squared test consistently outperforms the eigenvalue criterion. 

\begin{figure}[t]
	\def\svgwidth{0.8\columnwidth}
	\hspace*{4mm}
	\scalebox{1.1}{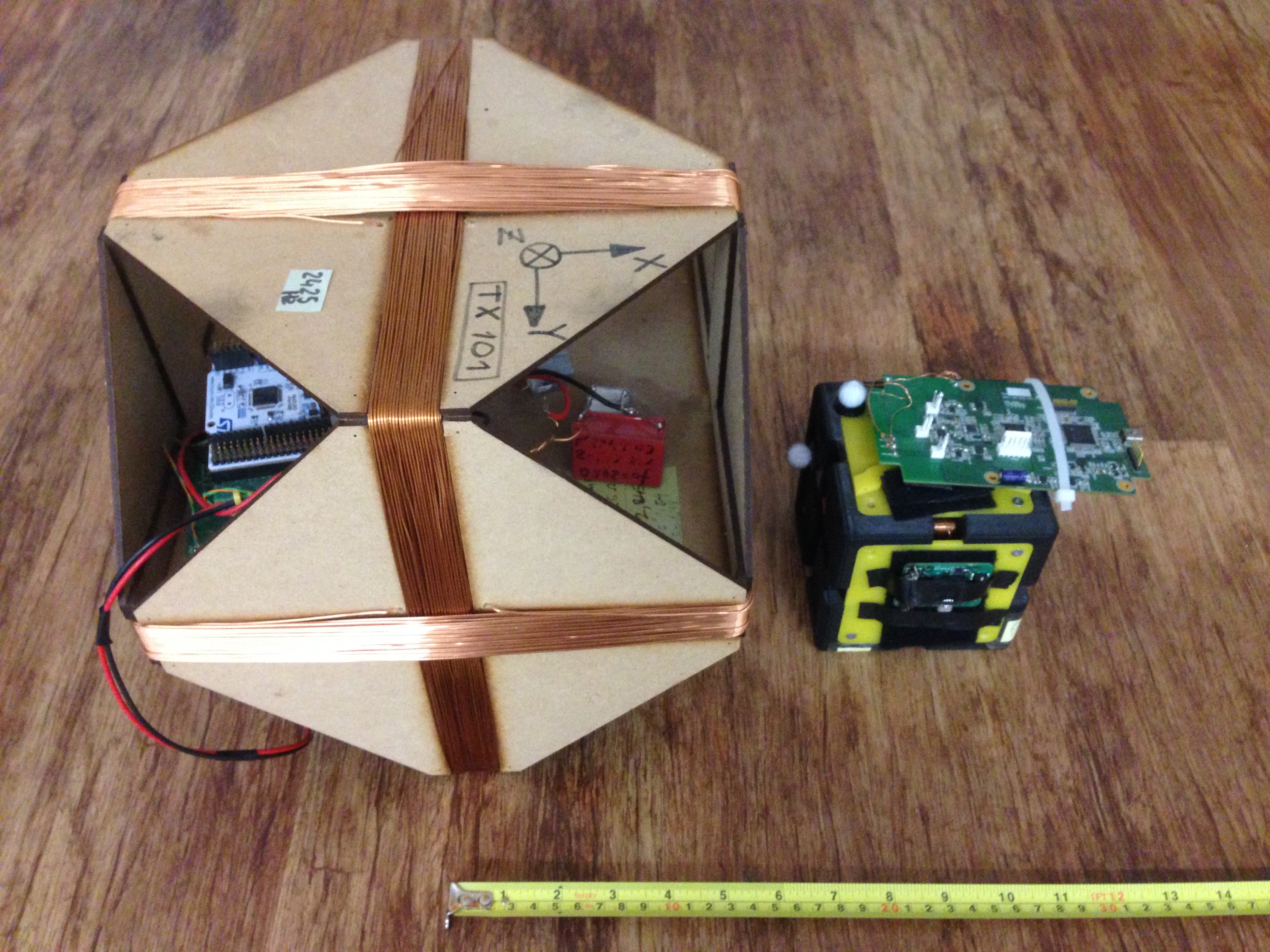}
	\vspace*{0mm}
	\caption{The transmitter (left) and receiver with an attached inertial measurements unit (right).}
	\label{fig_setup}
	\vspace*{-0mm}
\end{figure}

\begin{figure}[t]
	\vspace*{0mm}
	\hspace*{-1mm}
	\includegraphics{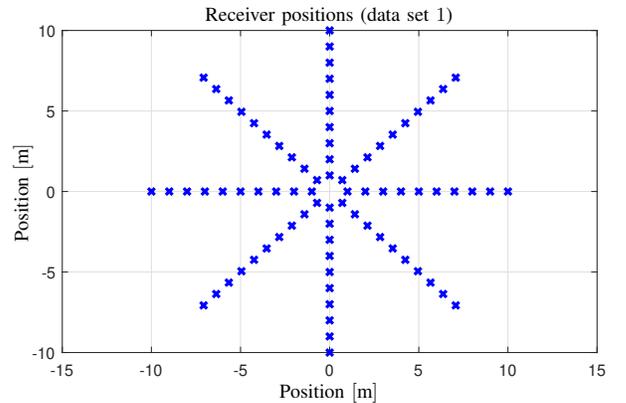}
	\vspace*{-6mm}
	\caption{Horizontal positions of receiver (relative to transmitter) in experiment.}
	\vspace*{0mm}
	\label{fig_experiments_true_positions_1}
\end{figure}

\section{experiments}

This section demonstrates the performance of the proposed estimator and fault detector on experimental data. The setup is illustrated in Fig. \ref{fig_setup}.

\begin{figure}[t]
	\hspace*{-1.5mm}
	\vspace*{3mm}
	{
		\includegraphics{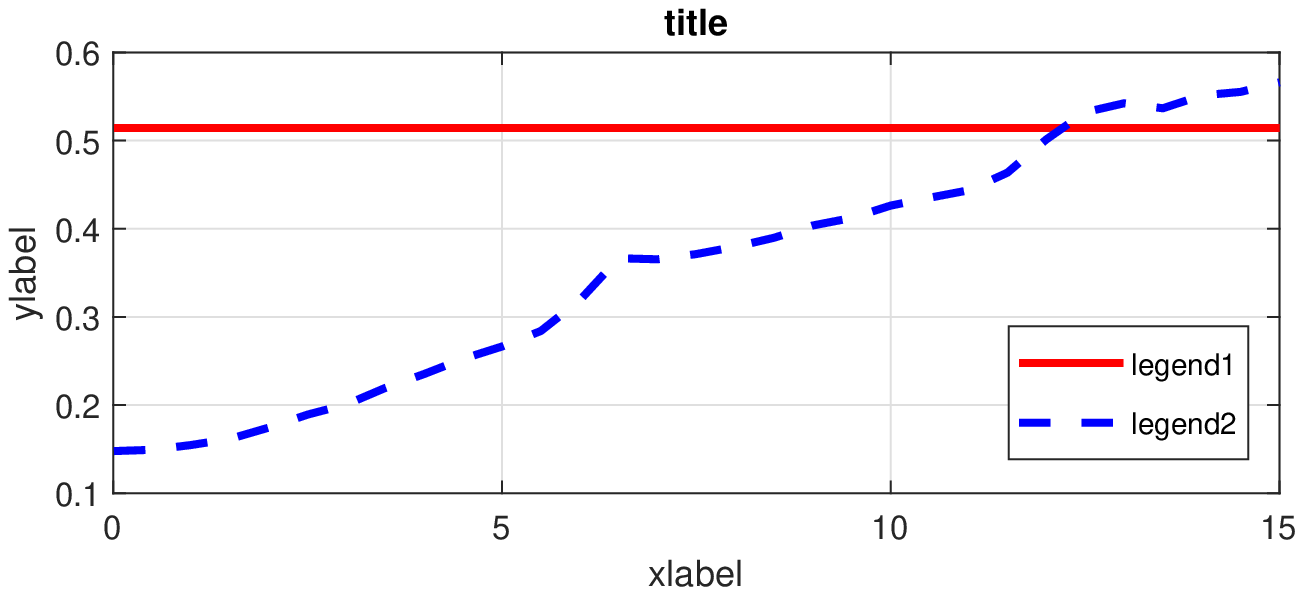}
		\vspace*{3mm}
	}
	\hspace*{-2.2mm}
	{
		\vspace*{-1.3mm}
		\includegraphics{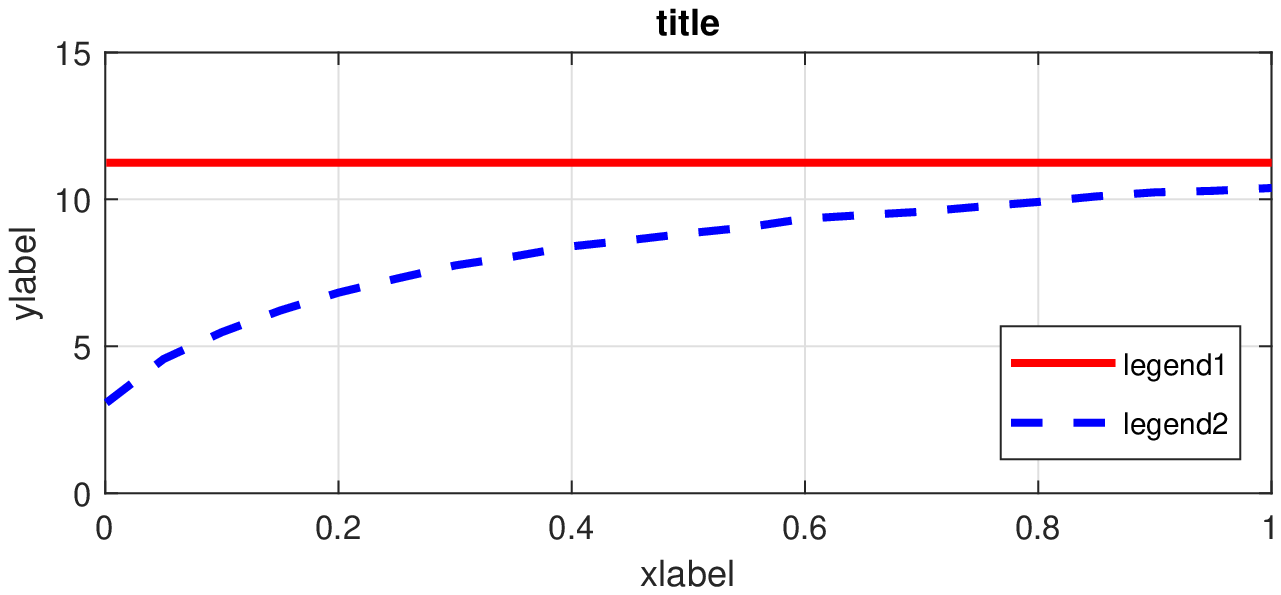}
		\vspace*{0mm}
	}
	\vspace*{-8mm}
	\caption{Accuracy of position and orientation estimates on experimental data. The data was collected at the receiver positions shown in Fig. \ref{fig_experiments_true_positions_1}. \label{fig_experiments_accuracy}}
	\vspace*{0mm}
\end{figure}

\subsection{Position and Orientation Estimation}

Data was collected outdoors from a receiver placed, in turn, at the 80 different positions displayed in Fig. \ref{fig_experiments_true_positions_1}. Along the vertical direction, the transmitter was placed $0.84\,[\text{m}]$ above the receiver. Both the receiver and the transmitter remained fixed during all data collections, and the performance evaluation used one packet (one set of MI measurements $\mathbf{y}\overset{_\Delta}{=}\{\mathbf{y}_k\}_{k=1}^N$) from each position. The roll and pitch angles were zero, while the yaw angle took on eight discrete values depending on the position of the receiver (the orientation of the receiver was adjusted so that a specific axis on the receiver always pointed towards the transmitter). The parameters $c$ and $\sigma$ were selected to match the recorded experimental data and remained fixed within each data set. Refer to \cite{Abrudan2015,Abrudan2016}, and \cite{Markham2012} for details on the experimental setup and the communication layer.

Fig. \ref{fig_experiments_accuracy} shows the results obtained with the same simulated priors as in Fig. \ref{fig_simulations_accuracy} but with experimental MI data and using $10^2$ simulations. As can be seen, the accuracy of the position estimates is clearly improved already with an orientation uncertainty of $10\,[^\circ]$, and the position error is about $0.15\,[\text{m}]$ with an orientation uncertainty of $1\,[^\circ]$. In real-world experiments, the errors of the MI signals do not follow a perfect Gaussian distribution, and hence, the MAP estimator may perform worse than the ML estimator.

\begin{figure}[t]
	\hspace*{-1.8mm}
	\vspace*{2.5mm}
	{
		\includegraphics{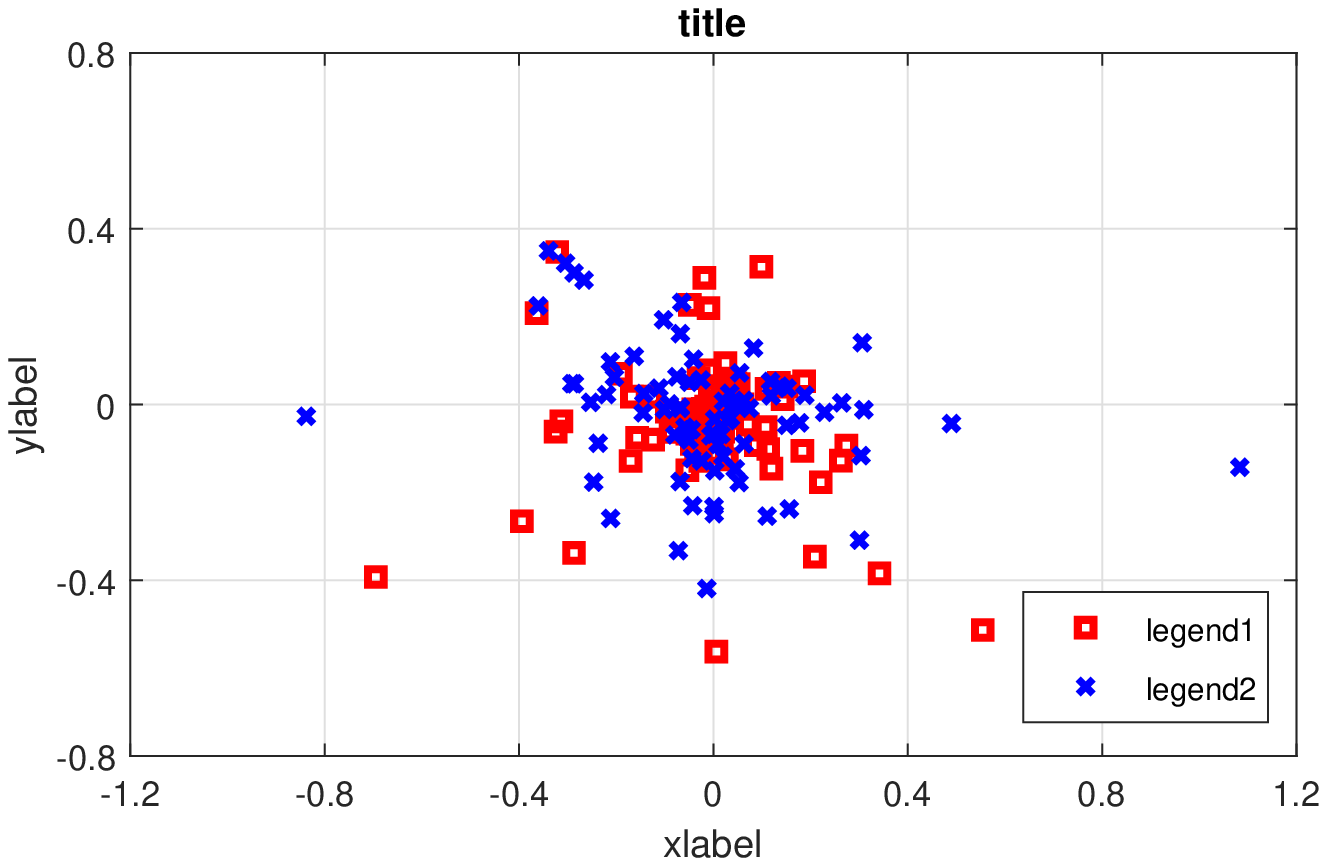}
		\vspace*{3mm}
	}
	\hspace*{-2.61mm}
	\vspace*{0mm}
	{
		\vspace*{-1.3mm}
		\includegraphics{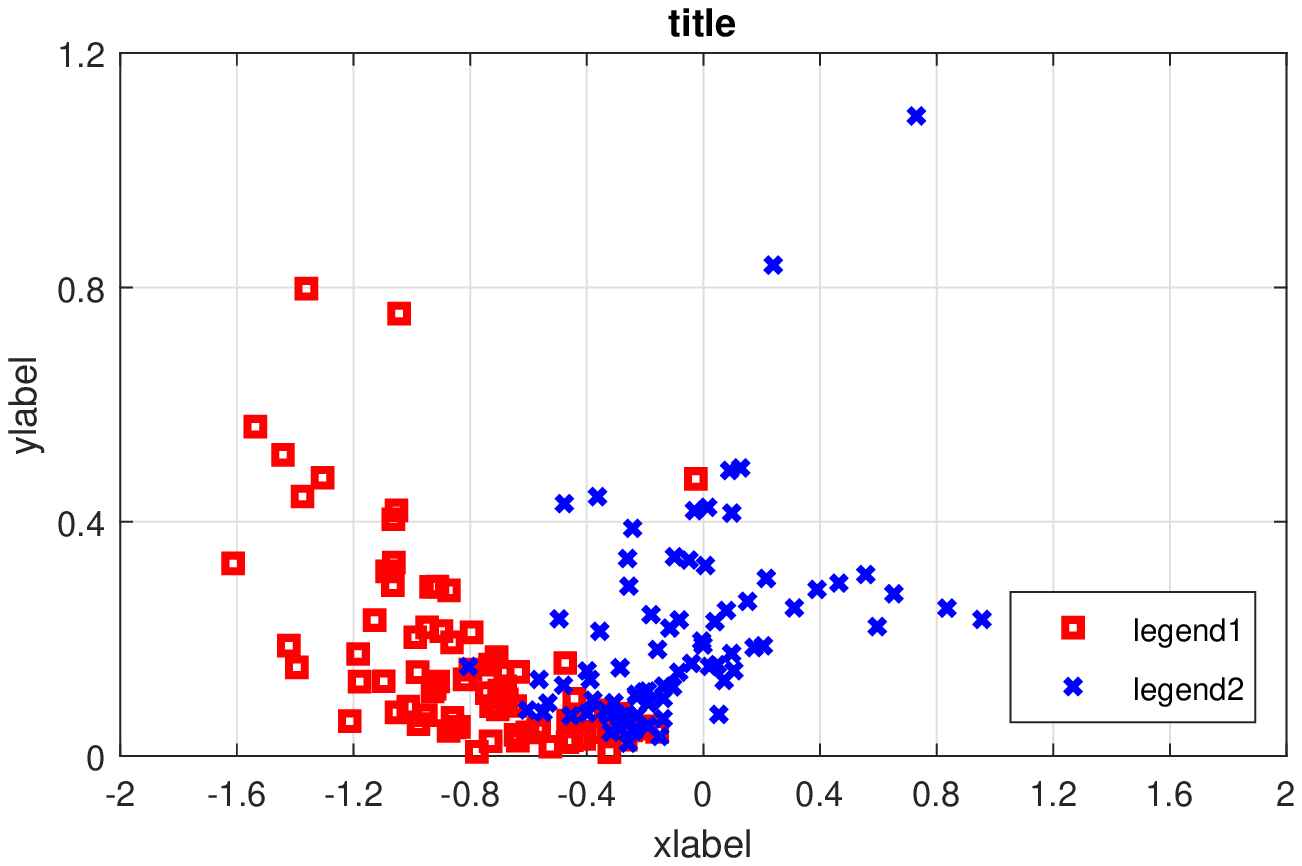}
	\vspace*{0mm}
	}
	\vspace*{-5mm}
	\caption{Example of position errors when simulating an orientation prior with an uncertainty of $5\,[^\circ]$. The data was collected at the receiver positions shown in Fig. \ref{fig_experiments_true_positions_1}. \label{fig_experiments_position_example}}
	\vspace*{0mm}
\end{figure}

\begin{figure}[t]
	\hspace*{-1.8mm}
	\vspace*{2.5mm}
	{
		\includegraphics{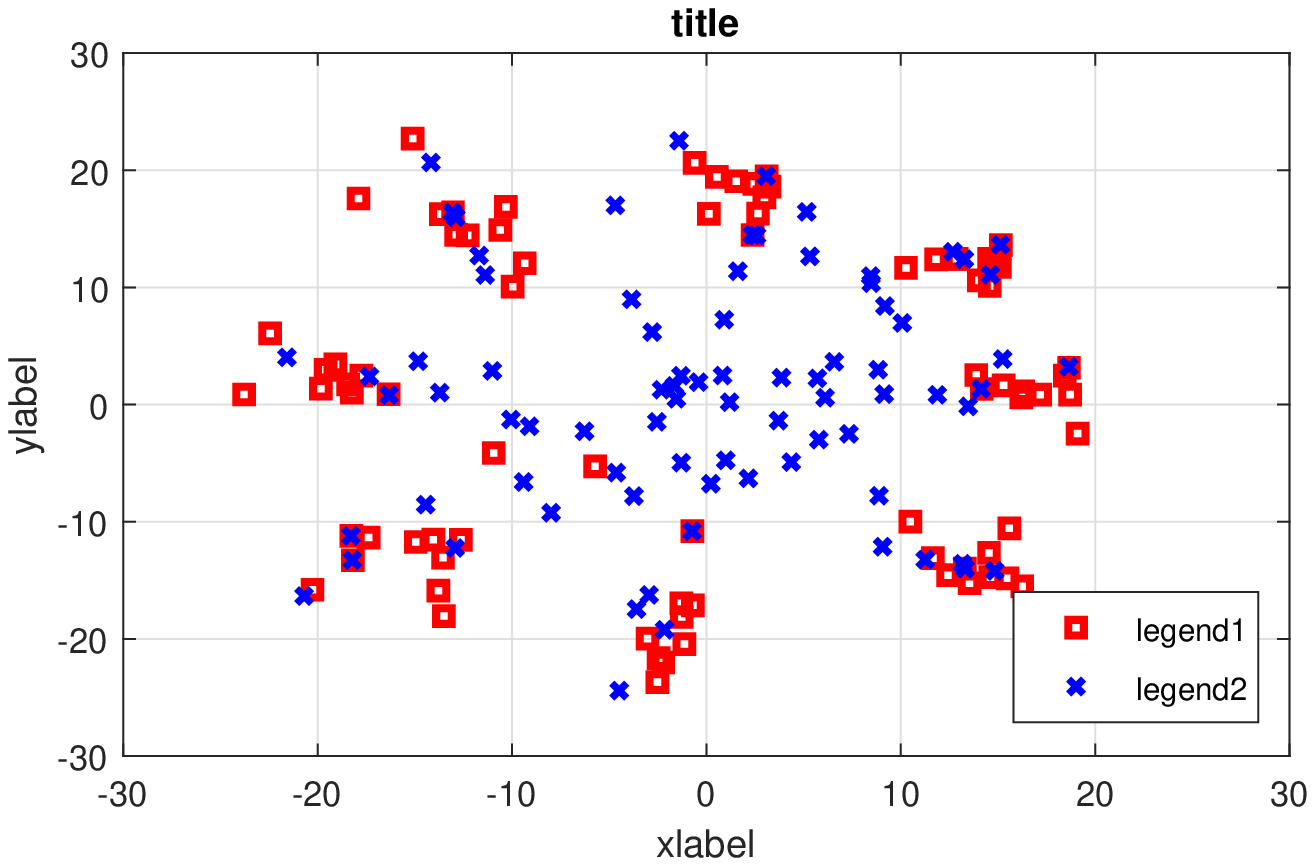}
		\vspace*{3mm}
	}
	\hspace*{-2.61mm}
	\vspace*{0mm}
	{
		\vspace*{-1.3mm}
		\includegraphics{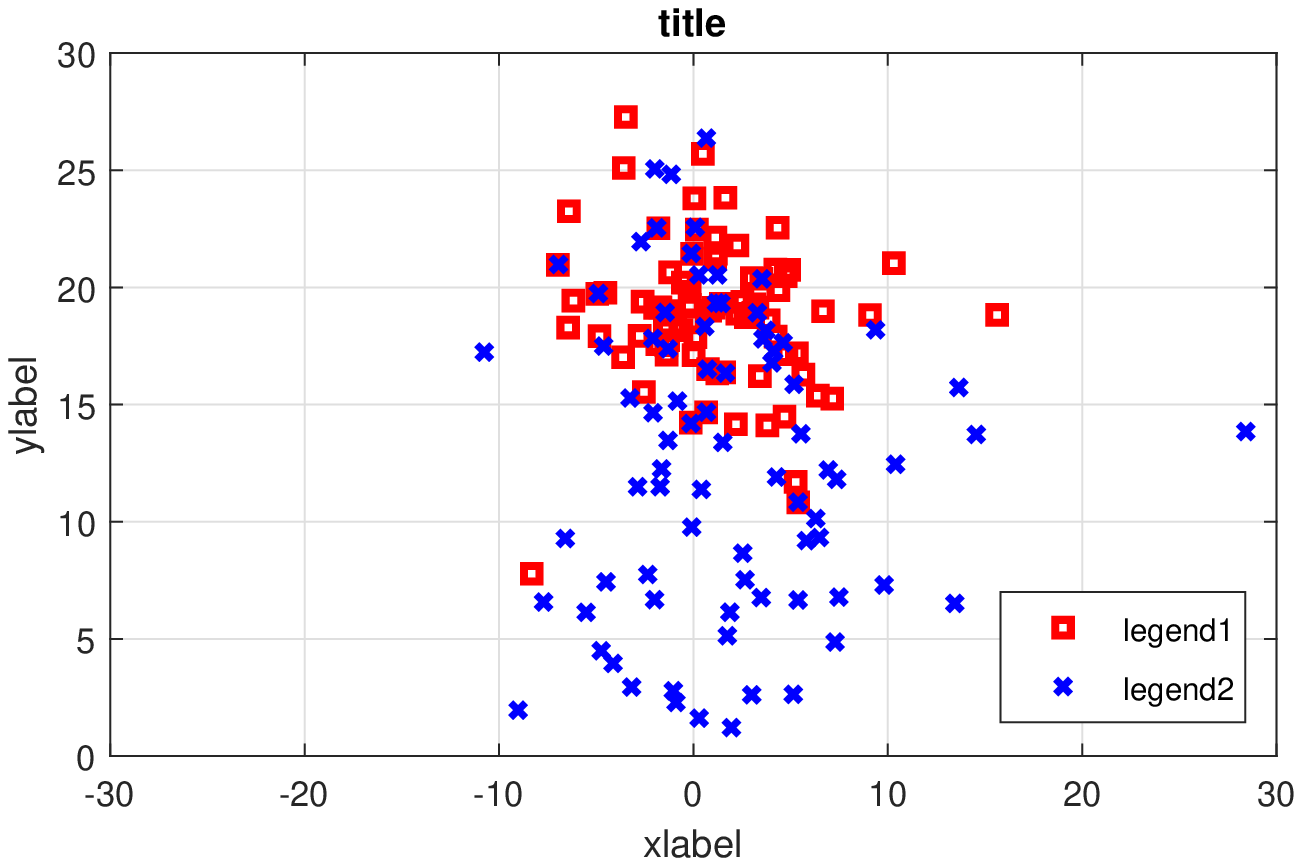}
		\vspace*{0mm}
	}
	\vspace*{-5mm}
	\caption{Example of orientation errors when simulating a position prior with an uncertainty of $0.5\,[\text{m}]$. The data was collected at the receiver positions shown in Fig. \ref{fig_experiments_true_positions_1}. \label{fig_experiments_orientation_example}}
	\vspace*{0mm}
\end{figure}

The position and orientation errors from an arbitrarily chosen simulation are shown in Figs. \ref{fig_experiments_position_example} and \ref{fig_experiments_orientation_example}, respectively. Two observations can be made. First, in correspondence with the analysis in Section \ref{section_CRB}, the ML position estimates are better in the horizontal plane than in the vertical direction (the RMSEs of the ML estimator were $0.16\,[\text{m}]$ and $0.17\,[\text{m}]$ in the two horizontal directions and $0.86\,[\text{m}]$ in the vertical direction). As described in \cite{Kypris2016}, this may also be a consequence of distortions. Likewise, in similarity with the results presented in \cite{Arumugam2012} and \cite{Arumugam2014}, the ML orientation estimates are better for the yaw angle than for the roll and pitch angles. Second, the performance improvement resulting from using the prior information is the greatest along those dimensions where the ML estimator exhibits its worst performance, i.e., along the vertical position dimension and for the roll and pitch angles. The same tendency was observed in \cite{Kypris2016} when comparing the ML estimator with an estimator that modeled distortions caused by metallic reinforcement bars. Last, note that the orientation errors are comparable to those presented in \cite{Arumugam2012}; the ML estimator produced median absolute errors of $12.55\,[^\circ]$ and $2.58\,[^\circ]$ for the pitch and yaw angles, respectively, whereas \cite{Arumugam2012} reported median absolute errors of $8.68\,[^\circ]$ and $4.08\,[^\circ]$ for the same angles.

\begin{figure}[t]
	\vspace*{0mm}
	\hspace*{-1mm}
	\includegraphics{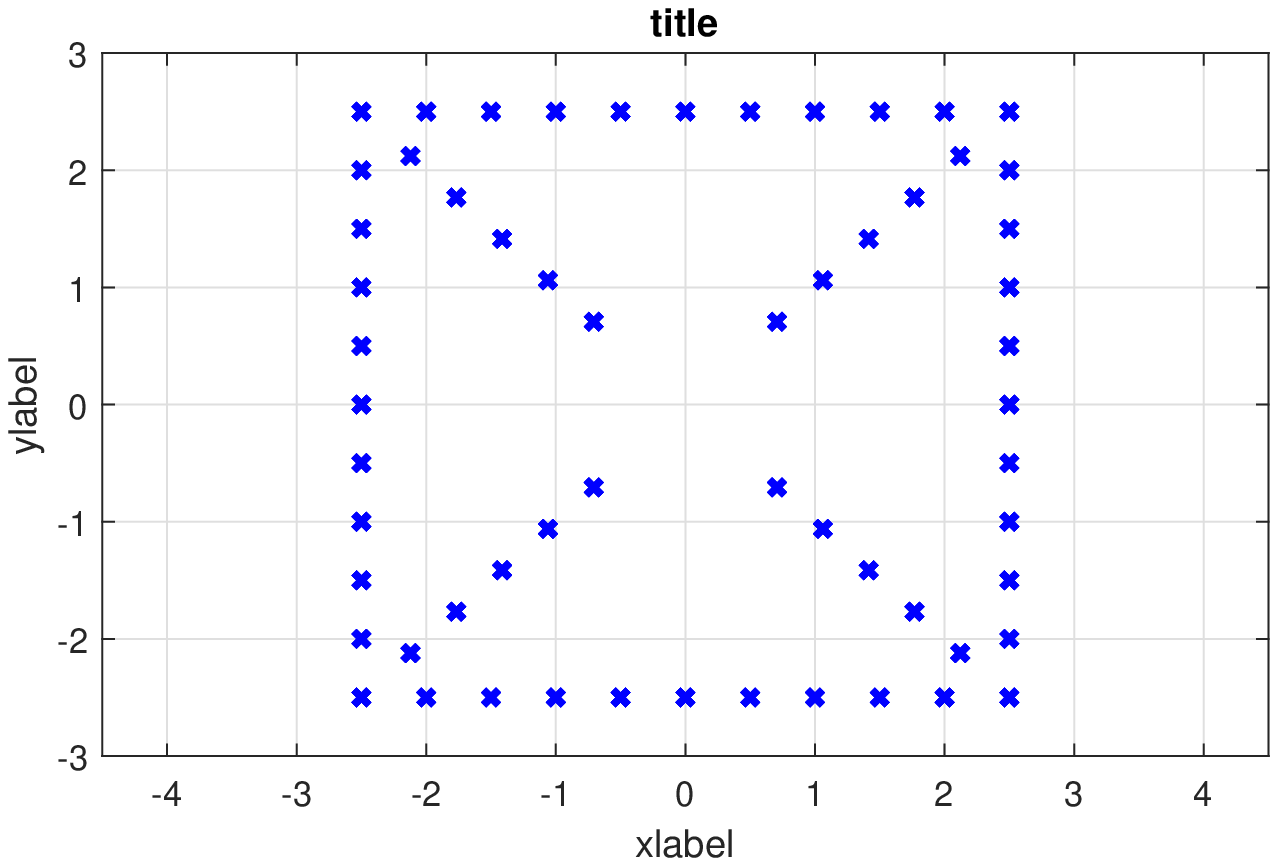}
	\vspace*{-6mm}
	\caption{Horizontal positions of receiver (relative to transmitter) in experiment.}
	\vspace*{0mm}
	\label{fig_experiments_true_positions_2}
\end{figure}

\begin{figure}[t]
	\vspace*{0mm}
	\hspace*{-1mm}
	\includegraphics{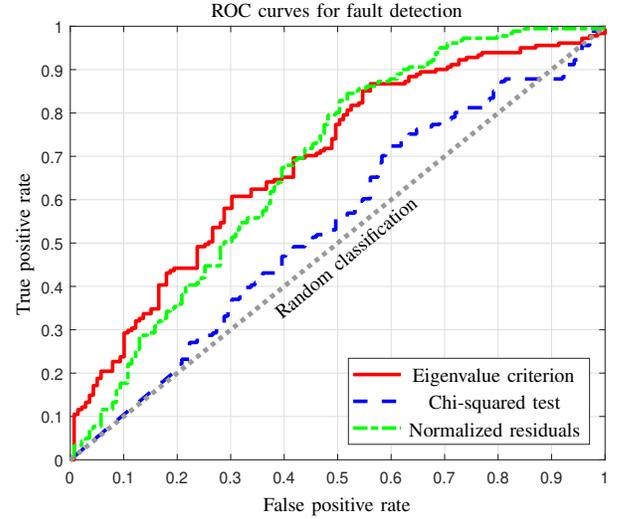}
	\vspace*{-6mm}
	\caption{Receiver operating characteristics on experimental data.}
	\vspace*{0mm}
	\label{fig_experiments_ROC}
\end{figure}

\subsection{Detection of Distortions}

Next, the fault detection methods are evaluated on experimental data. In addition to the outdoor data described in the previous subsection, we made use of data collected in a distorted indoor environment at the 60 positions illustrated in Fig. \ref{fig_experiments_true_positions_2}. In total, we used 240 indoor packets (four packets per position, collected at varying transmitter-receiver heights and orientations) and one packet from each of the 80 outdoor positions. In similarity with \cite{Abrudan2015}, position estimation errors larger than one meter were considered to be distorted. This corresponded to 159 indoor estimates and 22 outdoor estimates. Fig. \ref{fig_experiments_ROC} displays the ROC curves obtained from experimental data. In this case, the eigenvalue criterion performs better than the $\chi^2$ test. One possible explanation for this is that the magnitude of the measurement errors scale with the magnitude of the signal (something which is not captured by the measurement model \eqref{eq_meas_model}). To investigate this, we modified the chi-squared test by normalizing the sum of squared residuals $T(\mathbf{x})$. In this way, we compensate for variations in $T(\mathbf{x})$ that are associated with variations in the magnitude of the signal. Specifically, the detection was made based on the test statistic 
\begin{equation}
\label{eq_new_test_statistic}
T^{\hspace*{0.4mm}\diamond}(\mathbf{x})\overset{\Delta}{=}T(\mathbf{x})/\textstyle\sum_{k=1}^N\Vert\mathbf{y}_k\Vert^2.
\end{equation}
As illustrated in Fig. \ref{fig_experiments_ROC}, the performance of this detector is better than the chi-squared test and essentially equivalent to that of the eigenvalue criterion. This supports the notion that the failure of the chi-squared test on experimental data derives from limitations of the measurement model \eqref{eq_meas_model}.

\subsection{Fusion with Accelerometer Data}

MI and accelerometer data was collected outdoors at the 60 positions illustrated in Fig. \ref{fig_experiments_true_positions_2}. The performance evaluation used one packet from each position. Point estimates of the receiver's roll and pitch angles were obtained in the following way. The accelerometer measurements collected during the MI transmission were first averaged along each axis. Roll and pitch estimates were then computed from these averages based on standard formulas for accelerometer-based orientation estimation \cite{Groves2008}. These estimates were then incorporated into the MAP estimator as Gaussian priors with a standard deviation of $0.1\,[^\circ]$. The estimator did not make use of any prior information on the yaw angle. 

\begin{figure}[t]
	\hspace*{-1.5mm}
	\vspace*{0mm}
	\includegraphics{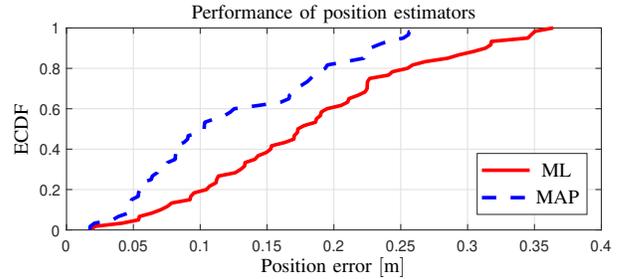}
	\hspace*{-1.4mm}
	\vspace*{-5mm}
	\caption{Empirical cumulative distribution functions of three-dimensional position errors produced by the ML estimator, using only MI data, and the MAP estimator, fusing MI and accelerometer data. \label{fig_experiments_fusion}}
	\vspace*{-3mm}
\end{figure}

Fig. \ref{fig_experiments_fusion} contrasts the performance of the ML estimator, using only MI data, and the MAP estimator, fusing MI and accelerometer data, by displaying the empirical cumulative distribution functions (ECDFs) of the magnitude of the three-dimensional position errors. The accelerometer measurements employed by the MAP estimator can be seen to enable a substantial performance enhancement, and reduce the median error almost by a factor of two. 

%

\section{summary}

This paper has examined magneto-inductive navigation from an estimation theoretical viewpoint. Specifically, we developed a MAP estimator for fusing MI signals with complementary information such as inertial measurements or motion constraints. Moreover, we proposed a chi-squared test for the detection of magnetic distortions and other measurement disturbances. Experimental results illustrated that the proposed MAP estimator can, by using prior information from example inertial sensors, yield a substantial performance improvement in comparison to a state-of-the-art ML estimator. Although the proposed detector demonstrated excellent performance on simulated data, distortion detection on field data remains a big challenge. This motivates further research into the mapping of magneto-inductive distortions using e.g., simultaneous localization and mapping algorithms.

\section*{appendix a}
\setcounter{subsection}{0}

This appendix outlines the method for 3-D MI position and orientation estimation presented in \cite{Abrudan2015} and \cite{Abrudan2016}. 

\subsection{Rotation Stabilization}

Generally, the navigation state changes during transmission. Hence, the measurement equation \eqref{eq_meas_model} can be modified as
\begin{equation}
\mathbf{y}_k=\mathbf{h}_{\mathbf{m}_k}(\mathbf{x}_k)+\mathbf{e}_k
\end{equation}
where $\mathbf{x}_k$ denotes the navigation state at sampling instance $k$. Changes in the orientation $\boldsymbol{\psi}$ can be estimated using gyroscope measurements. This gives us the orientations $\{\boldsymbol{\phi}_k\}_{k=2}^N$, where $\boldsymbol{\phi}_k$ represents the change in orientation between sampling instances $k$ and $k-1$. The estimated change in orientation is used with the intention of eliminating the time-dependence from the dipole model. Hence, we define 
\begin{equation}
\mathbf{z}_k\overset{\Delta}{=}\mathbf{R}(\boldsymbol{\phi}_k)^\intercal\hspace*{0.2mm}\mathbf{y}_k.
\end{equation}
Assuming that the estimates $\{\boldsymbol{\phi}_k\}_{k=2}^N$ are perfect and that the position $\mathbf{r}$ remains constant during the transmission, we have 
\begin{equation}
\mathbf{z}_k=\mathbf{h}_{\mathbf{m}_k}(\mathbf{x}_1)+\mathbf{w}_k
\end{equation}
for $k=1,\dots,N$, where $\mathbf{w}_k\overset{_\Delta}{=}\mathbf{R}(\boldsymbol{\phi}_k)^\intercal\mathbf{e}_k$. Thus, in this case, the time-dependence is eliminated from the measurement function and we have obtained a measurement equation of the same form as in \eqref{eq_meas_model}. Changes in the position $\mathbf{r}$ during transmission could potentially be compensated for by incorporating accelerometers. However, as far as the authors know, this has not been demonstrated in any published work. 

\subsection{Channel Matrix Estimation} 

The ML estimate $\widehat{\mathbf{S}}$ of the channel matrix $\mathbf{S}$ is computed based on linear least-squares. The channel matrix is defined as $\mathbf{S}\overset{_\Delta}{=}[\hspace*{0.2mm}\mathbf{h}_{\mathbf{m}_1}(\mathbf{x})\;\mathbf{h}_{\mathbf{m}_2}(\mathbf{x})\;\mathbf{h}_{\mathbf{m}_3}(\mathbf{x})\hspace*{0.2mm}]$ where 
$\mathbf{m}_{1}=[\hspace*{0.2mm}1\,0\,0\hspace*{0.2mm}]^\intercal$, $\mathbf{m}_{2}=[\hspace*{0.2mm}0\,1\,0\hspace*{0.2mm}]^\intercal$, and $\mathbf{m}_{3}=[\hspace*{0.2mm}0\,0\,1\hspace*{0.2mm}]^\intercal$. However, note that the constraints on $\mathbf{S}$ imposed by the dipole model are not enforced when computing $\widehat{\mathbf{S}}$.

\subsection{Position Estimation}

The ML estimate of the range $\Vert\mathbf{r}\Vert$ is first computed as 
\begin{equation}
\Vert\hat{\mathbf{r}}\Vert = r_0 10^{(\rho_0-\rho)/60}
\end{equation}
where the received signal strength indicator (RSSI) is $\rho\overset{_\Delta}{=}20\log_{10}\Vert\widehat{\mathbf{S}}\Vert_{\text{F}}$, $\Vert\Vert_{\text{F}}$ denotes the Frobenius norm, and $\rho_0$ is the RSSI measured at some reference distance $r_0$\footnote{$\rho_0$ could also be computed directly from the model if $c$ is known.}. Further, the position estimate is computed as
\begin{equation}
\hat{\mathbf{r}}=\Vert\hat{\mathbf{r}}\Vert\mathbf{u}_{\text{max}}
\end{equation} 
where $\mathbf{u}_{\text{max}}$ is the eigenvector associated with the maximal eigenvalue of the Gramian matrix $\widehat{\mathbf{S}}^\intercal \widehat{\mathbf{S}}$. Refer to \cite{Abrudan2015} for motivating details.

\subsection{Orientation Estimation}

The orientation is estimated as 
\begin{equation}
\mathbf{R}(\widehat{\boldsymbol{\psi}})=\frac{\text{pf}(\mathbf{U}\widehat{\mathbf{S}}^\intercal)}{\text{sign}(\det(\text{pf}(\mathbf{U}\widehat{\mathbf{S}}^\intercal)))}
\end{equation}
where $\mathbf{U}\overset{_\Delta}{=}3\mathbf{u}_{\text{max}}\mathbf{u}_{\text{max}}^\intercal-\mathbf{I}_3$ and $\text{pf}(\cdot)$ denotes the orthogonal polar factor \cite{Higham1989}. Refer to \cite{Abrudan2016} for motivating details.

\section*{appendix b}

Since 
\begin{equation}
\mathbf{S}^\intercal\mathbf{S}=\frac{c^2}{\Vert\mathbf{r}\Vert^6}\bigg(\frac{3\mathbf{r}\mathbf{r}^\intercal}{\Vert\mathbf{r}\Vert^2}+\mathbf{I}_3\bigg),
\end{equation}
the eigenvalues of $\mathbf{S}^\intercal\mathbf{S}$ are $c^2/\Vert\mathbf{r}\Vert^6\cdot[4\,1\,1]^\intercal$. Further, after normalizing the eigenvalues by dividing them with their average value $2c^2/\Vert\mathbf{r}\Vert^6$, we obtain 
\begin{equation}
\boldsymbol{\lambda}\overset{_\Delta}{=}[\hspace*{0.2mm}2\;\;1/2\;\;1/2\hspace*{0.2mm}]^\intercal.
\end{equation}
Thus, the presence of magnetic distortions may be detected by thresholding the eigenvalue criterion \cite{Abrudan2015}
\begin{equation}
J(\boldsymbol{\lambda})=\Vert\widehat{\boldsymbol{\lambda}}-\boldsymbol{\lambda}\Vert.
\end{equation}
Here, $\widehat{\boldsymbol{\lambda}}$ is the eigenvalue vector of
$\widehat{\mathbf{S}}^\intercal\widehat{\mathbf{S}}$, with the elements sorted in descending order and divided by their average value. 

\section*{appendix c}

This appendix derives the Fisher information in \eqref{eq_fish_inf}. The CRB for $\hat{\mathbf{x}}$ is\footnote{Generally, the use of CRBs (which apply to flat Euclidean spaces) to analyze the errors of Euler angles (which are circular parameters) is sensible when the bounds on the orientation angles are less than $10\,[^\circ]$ \cite{Gerwe2003}.}
\begin{equation}
\text{Cov}(\hat{\mathbf{x}})\succeq \boldsymbol{\mathcal{I}}_\mathbf{x}^{-1}
\end{equation}
where we have used $\mathbf{A}\succeq\mathbf{B}$ to denote that $\mathbf{A}-\mathbf{B}$ is positive semidefinite and where the Fisher information matrix is\footnote{The information matrix associated with the corresponding Bayesian CRB is $\boldsymbol{\mathcal{I}}_{\text{data}}+\boldsymbol{\mathcal{I}}_{\text{prior}}$ where $\boldsymbol{\mathcal{I}}_{\text{data}}$ is the expectation of $\boldsymbol{\mathcal{I}}_\mathbf{x}$ with respect to the prior and $\boldsymbol{\mathcal{I}}_{\text{prior}}$ is the information provided by the prior \cite{VanTrees1968,Wahlstrom2018}.} \cite{Kay1993}
\begin{equation}
\boldsymbol{\mathcal{I}}_\mathbf{x}=\sum_{k=1}^N(\partial_\mathbf{x}\mathbf{h}_{\mathbf{m}_k}(\mathbf{x}))^\intercal\mathbf{P}^{-1}\partial_\mathbf{x}\mathbf{h}_{\mathbf{m}_k}(\mathbf{x}).
\end{equation}
Here, $\partial_\mathbf{x}\mathbf{h}_{\mathbf{m}_k}(\mathbf{x})$ is the Jacobian of $\mathbf{h}_{\mathbf{m}_k}(\mathbf{x})$ with respect to $\mathbf{x}$. The Fisher information $\boldsymbol{\mathcal{I}}_\mathbf{x}$ can easily be computed based on the dipole field model. However, for our analytical results, we will focus on the position CRB under the assumptions specified in Section \ref{section_CRB}.

Consider the case when the orientation $\boldsymbol{\psi}$ is known. The CRB for the position estimates can then be written as 
\begin{equation}
\text{Cov}(\hat{\mathbf{r}})\succeq \boldsymbol{\mathcal{I}}_\mathbf{r}^{-1}
\end{equation}
where the Fisher information matrix is
\begin{equation}
\boldsymbol{\mathcal{I}}_\mathbf{r}=\sum_{k=1}^N(\partial_\mathbf{r}\mathbf{h}_{\mathbf{m}_k}(\mathbf{x}))^\intercal\mathbf{P}^{-1}\partial_\mathbf{r}\mathbf{h}_{\mathbf{m}_k}(\mathbf{x}).
\end{equation}
Here, it holds that \cite{Wahlstrom2014}
\begin{align}
\partial_\mathbf{r}\mathbf{h}_{\mathbf{m}}(\mathbf{x})&=c\cdot \mathbf{R}(\boldsymbol{\psi})\frac{3}{\Vert\mathbf{r}\Vert^5}\big((\mathbf{r}\boldsymbol{\cdot}\mathbf{m})\cdot\mathbf{I}_3\\
&\hspace*{3.2mm}+\mathbf{r}\mathbf{m}^\intercal+\mathbf{m}\mathbf{r}^\intercal-5\frac{(\mathbf{r}\boldsymbol{\cdot}\mathbf{m})}{(\mathbf{r}\boldsymbol{\cdot}\mathbf{r})}\mathbf{r}\mathbf{r}^\intercal\bigg)\nonumber.
\end{align}
Further, using that $\mathbf{R}(\boldsymbol{\psi})^\intercal\mathbf{R}(\boldsymbol{\psi})= \mathbf{I}_3$ and $\mathbf{P}=\sigma^2\cdot\mathbf{I}_9$,
straightforward but tedious calculations give that the FIM is
\begin{equation}
\boldsymbol{\mathcal{I}}_\mathbf{r}=\sum_{k=1}^N\mathbf{g}(\mathbf{r},\mathbf{m}_k)
\end{equation}
where
\begin{align}
\mathbf{g}(\mathbf{r},\mathbf{m})&\overset{_\Delta}{=}\frac{9Nc^2}{\sigma^2\Vert\mathbf{r}\Vert^{10}}\big((\mathbf{r}\boldsymbol{\cdot}\mathbf{m})^2\cdot\mathbf{I}_3+(\mathbf{r}\boldsymbol{\cdot}\mathbf{r})\mathbf{m}\mathbf{m}^\intercal\nonumber\\
&\hspace*{3.2mm}-2(\mathbf{r}\boldsymbol{\cdot}\mathbf{m})(\mathbf{r}\mathbf{m}^\intercal+\mathbf{m}\mathbf{r}^\intercal)\\
&\hspace*{3.2mm}+\Big(5\frac{(\mathbf{r}\boldsymbol{\cdot}\mathbf{m})^2}{(\mathbf{r}\boldsymbol{\cdot}\mathbf{r})}+(\mathbf{m}\boldsymbol{\cdot}\mathbf{m})\Big)\mathbf{r}\mathbf{r}^\intercal\bigg).\nonumber
\end{align}
Now, assuming that $N$ is equal to a multiple of 3 while $\mathbf{m}_{1+3k}=[\hspace*{0.2mm}m\,0\,0\hspace*{0.2mm}]^\intercal$, $\mathbf{m}_{2+3k}=[\hspace*{0.2mm}0\,m\,0\hspace*{0.2mm}]^\intercal$, and $\mathbf{m}_{3+3k}=[\hspace*{0.2mm}0\,0\,m\hspace*{0.2mm}]^\intercal$ for all $k$, the $(i,i)$th element of $\boldsymbol{\mathcal{I}}_\mathbf{r}$ is
\begin{equation}
\mathcal{I}_{r_i}=\frac{6Nc^2m^2}{\sigma^2\Vert\mathbf{r}\Vert^8}\bigg(1+2\frac{[\mathbf{r}]_i^2}{\Vert\mathbf{r}\Vert^2}\bigg)
\end{equation}
which completes the derivation of \eqref{eq_fish_inf}.

\section*{appendix d}

This appendix derives the Fisher information in \eqref{eq_fish_inf_range}. The CRB for the range estimates is 
\begin{equation}
\text{Var}(\Vert\hat{\mathbf{r}}\Vert)\ge \mathcal{I}_{\Vert\mathbf{r}\Vert}^{-1}
\end{equation}
where the Fisher information matrix is \cite{Kay1993}
\begin{equation}
\mathcal{I}_{\Vert\mathbf{r}\Vert}=\sum_{k=1}^N (\partial_{\Vert\mathbf{r}\Vert}\mathbf{h}_{\mathbf{m}_k}(\mathbf{x}))^\intercal\mathbf{P}^{-1}\partial_{\Vert\mathbf{r}\Vert}\mathbf{h}_{\mathbf{m}_k}(\mathbf{x}).
\end{equation}
Further, noting that the term $\mathbf{r}\mathbf{r}^\intercal/\Vert\mathbf{r}\Vert^2$ is range-independent, we have that
\begin{equation}
\partial_{\Vert\mathbf{r}\Vert}\mathbf{h}_{\mathbf{m}}(\mathbf{x})\overset{_\Delta}{=}-3c\cdot\frac{\mathbf{R}(\boldsymbol{\psi})}{\Vert\mathbf{r}\Vert^4}\bigg(\frac{3\mathbf{r}\mathbf{r}^\intercal}{\Vert\mathbf{r}\Vert^2}-\mathbf{I}_3\bigg)\mathbf{m}.
\end{equation}
Now, we assume that $N$ is equal to a multiple of 3 while $\mathbf{m}_{1+3k}=[\hspace*{0.2mm}m\,0\,0\hspace*{0.2mm}]^\intercal$, $\mathbf{m}_{2+3k}=[\hspace*{0.2mm}0\,m\,0\hspace*{0.2mm}]^\intercal$, and $\mathbf{m}_{3+3k}=[\hspace*{0.2mm}0\,0\,m\hspace*{0.2mm}]^\intercal$ for all $k$, and use that $\mathbf{R}(\boldsymbol{\psi})^\intercal\mathbf{R}(\boldsymbol{\psi})= \mathbf{I}_3$ and $\mathbf{P}=\sigma^2\cdot\mathbf{I}_9$. Straightforward calculations then give that 
\begin{align}
\begin{split}
\mathcal{I}_{\Vert\mathbf{r}\Vert}&=\frac{3Nc^2m^2}{\sigma^2\Vert\mathbf{r}\Vert^8}\sum_{i=1}^3\bigg(1+3\frac{[\mathbf{r}]_i^2}{\Vert\mathbf{r}\Vert^2}\bigg)\\
&=\frac{18Nc^2m^2}{\sigma^2\Vert\mathbf{r}\Vert^8}
\end{split}
\end{align}
which completes the derivation of \eqref{eq_fish_inf_range}.


\ifCLASSOPTIONcaptionsoff
  \newpage
\fi



\bibliographystyle{IEEEtran}
\bibliography{refs}

\end{document}

%% file: illustration.pdf_tex
\begingroup%
  \makeatletter%
  \providecommand\color[2][]{%
    \errmessage{(Inkscape) Color is used for the text in Inkscape, but the package 'color.sty' is not loaded}%
    \renewcommand\color[2][]{}%
  }%
  \providecommand\transparent[1]{%
    \errmessage{(Inkscape) Transparency is used (non-zero) for the text in Inkscape, but the package 'transparent.sty' is not loaded}%
    \renewcommand\transparent[1]{}%
  }%
  \providecommand\rotatebox[2]{#2}%
  \newcommand*\fsize{\dimexpr\f@size pt\relax}%
  \newcommand*\lineheight[1]{\fontsize{\fsize}{#1\fsize}\selectfont}%
  \ifx\svgwidth\undefined%
    \setlength{\unitlength}{1275.59055118bp}%
    \ifx\svgscale\undefined%
      \relax%
    \else%
      \setlength{\unitlength}{\unitlength * \real{\svgscale}}%
    \fi%
  \else%
    \setlength{\unitlength}{\svgwidth}%
  \fi%
  \global\let\svgwidth\undefined%
  \global\let\svgscale\undefined%
  \makeatother%
  \begin{picture}(1,0.66666667)%
    \lineheight{1}%
    \setlength\tabcolsep{0pt}%
    \put(0,0){\includegraphics[width=\unitlength,page=1]{illustration.pdf}}%
    \put(0.68485928,0.5409284){\color[rgb]{0,0,0}\makebox(0,0)[lt]{\begin{minipage}{0.06522469\unitlength}\raggedright \end{minipage}}}%
    \put(0.77338484,0.54269504){\color[rgb]{0,0,0}\makebox(0,0)[lt]{\lineheight{1.25}\smash{\begin{tabular}[t]{l}$\boldsymbol{\psi}$\end{tabular}}}}%
    \put(0,0){\includegraphics[width=\unitlength,page=2]{illustration.pdf}}%
    \put(0.34065442,0.12625445){\color[rgb]{0,0,0}\makebox(0,0)[lt]{\lineheight{1.25}\smash{\begin{tabular}[t]{l}$\mathbf{r}$\end{tabular}}}}%
    \put(0.06195998,0.60603223){\color[rgb]{0,0,0}\makebox(0,0)[lt]{\lineheight{1.25}\smash{\begin{tabular}[t]{l}Transmitter\end{tabular}}}}%
    \put(0.63080864,0.0136275){\color[rgb]{0,0,0}\makebox(0,0)[lt]{\lineheight{1.25}\smash{\begin{tabular}[t]{l}Receiver\end{tabular}}}}%
    \put(0,0){\includegraphics[width=\unitlength,page=3]{illustration.pdf}}%
    \put(0.30625315,0.38874784){\color[rgb]{0,0,0}\makebox(0,0)[lt]{\lineheight{1.25}\smash{\begin{tabular}[t]{l}Magnetic\end{tabular}}}}%
    \put(0.34656632,0.32240353){\color[rgb]{0,0,0}\makebox(0,0)[lt]{\lineheight{1.25}\smash{\begin{tabular}[t]{l}signal\end{tabular}}}}%
    \put(0,0){\includegraphics[width=\unitlength,page=4]{illustration.pdf}}%
  \end{picture}%
\endgroup%

%% file: setup.pdf_tex
\begingroup%
  \makeatletter%
  \providecommand\color[2][]{%
    \errmessage{(Inkscape) Color is used for the text in Inkscape, but the package 'color.sty' is not loaded}%
    \renewcommand\color[2][]{}%
  }%
  \providecommand\transparent[1]{%
    \errmessage{(Inkscape) Transparency is used (non-zero) for the text in Inkscape, but the package 'transparent.sty' is not loaded}%
    \renewcommand\transparent[1]{}%
  }%
  \providecommand\rotatebox[2]{#2}%
  \newcommand*\fsize{\dimexpr\f@size pt\relax}%
  \newcommand*\lineheight[1]{\fontsize{\fsize}{#1\fsize}\selectfont}%
  \ifx\svgwidth\undefined%
    \setlength{\unitlength}{598.11023622bp}%
    \ifx\svgscale\undefined%
      \relax%
    \else%
      \setlength{\unitlength}{\unitlength * \real{\svgscale}}%
    \fi%
  \else%
    \setlength{\unitlength}{\svgwidth}%
  \fi%
  \global\let\svgwidth\undefined%
  \global\let\svgscale\undefined%
  \makeatother%
  \begin{picture}(1,0.74881517)%
    \lineheight{1}%
    \setlength\tabcolsep{0pt}%
    \put(0,0){\includegraphics[width=\unitlength,page=1]{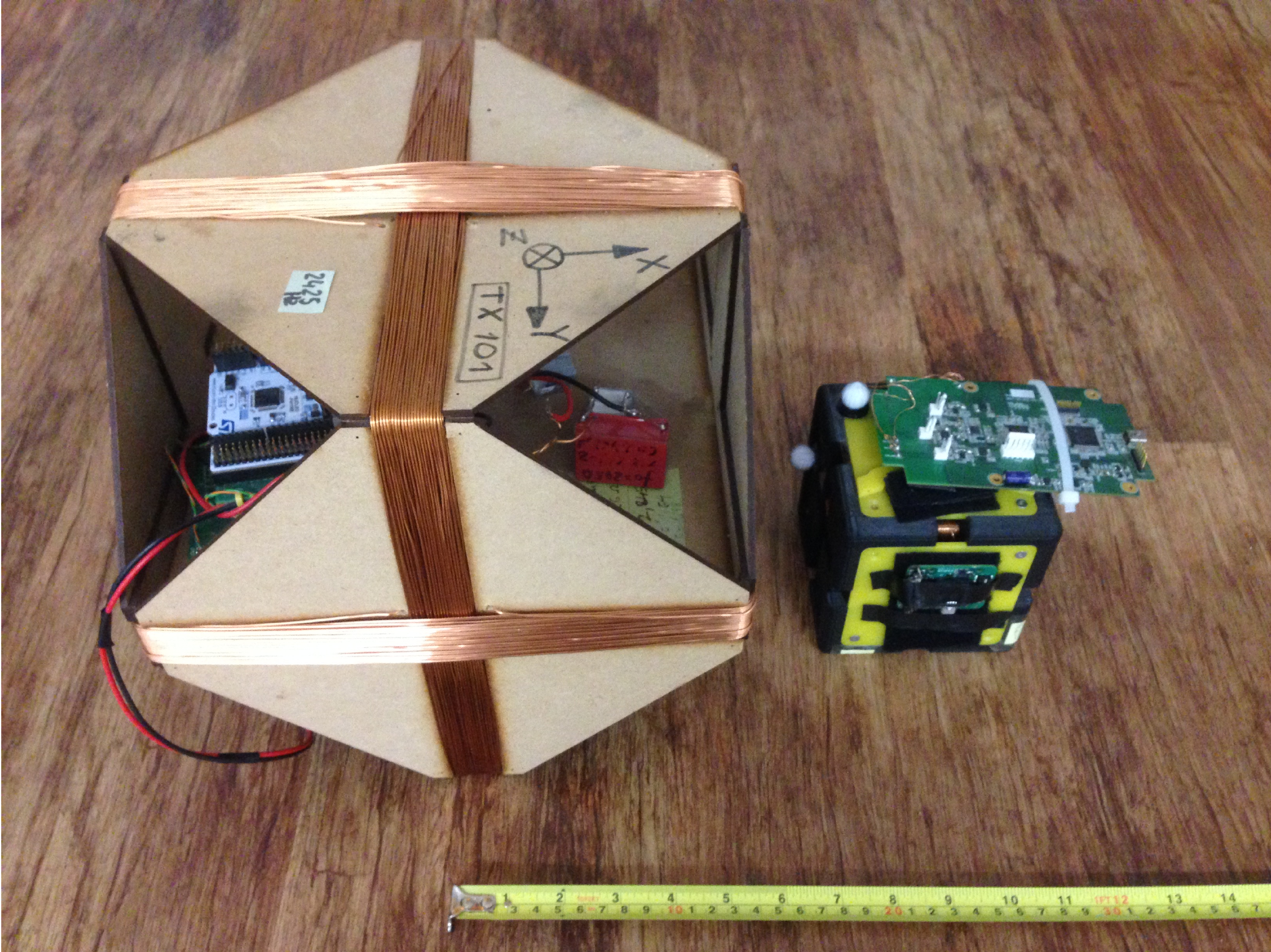}}%
  \end{picture}%
\endgroup%